\documentclass[oneside,english]{elsart}
\usepackage[T1]{fontenc}
\usepackage[latin1]{inputenc}
\usepackage{graphicx}
\usepackage{amssymb}

\makeatletter

\providecommand{\tabularnewline}{\\}

\usepackage{graphicx}
\usepackage{bm}
\newcommand{\bu}{{\bm u}}

\usepackage{babel}
\makeatother
\begin{document}
\begin{frontmatter}

\title{Sawtooth disruptions and limit cycle oscillations}

\author{Madhurjya P. Bora and Dipak Sarmah}

\address{Physics Department, Gauhati University, Guwahati 781014, India.}

\ead{mpbguw5@dataone.in}

\begin{abstract}
A minimal (low-dimensional) dynamical model of the sawtooth oscillations
is presented. It is assumed that the sawtooth is triggered by a thermal
instability which causes the plasma temperature in the central part
of the plasma to drop suddenly, leading to the sawtooth crash. It
is shown that this model possesses an isolated limit cycle which exhibits
relaxation oscillation, in the appropriate parameter regime, which
is the typical characteristics of sawtooth oscillations. It is further
shown that the invariant manifold of the model is actually the slow
manifold of the relaxation oscillation.
\end{abstract}
\begin{keyword}
Tokamak \sep Sawtooth oscillation \sep Limit cycle \sep Nonlinear dynamics \sep Dynamical model

\PACS 52.55.F \sep 02.60.Cb \sep 02.60.Lj
\end{keyword}
\end{frontmatter}

\section{Introduction}

Sawtooth oscillations \cite{goeler}, commonly observed in current
carrying, magnetically confined plasmas, are believed to be the result
of resistive internal kink mode i.e. $(m=1,n=1)$ oscillation. These
oscillations are characterized by a relatively slow rise of the electron
temperature in the central region of the plasma column followed by
a rapid drop (the crash). This is a typical nature of a \textit{stick-slip}
or \textit{relaxation} oscillation \cite{strogatz}, where the stress
is slowly built up and then suddenly released after a certain threshold,
observed in many other dynamical systems e.g. Portevin-Le Chatelier
effect \cite{portevin}, a matter of interest in material science. 

In this work, we propose a \emph{minimal} (low-dimensional) dynamical
model for sawtooth oscillation in tokamaks, based on a transport catastrophe
due to a thermal instability \cite{coppi1}. Low-dimensional or low-order
dynamical system i.e. a system of coupled ordinary differential equations,
has several advantages over a detailed physical model of the actual
system in the understanding of the global behaviour of the system.
These models become powerful as they are supported by well developed
mathematical theories which can be used to gain insight into the qualitative
behaviour of the system such as bifurcation and stability \cite{holmes}.
Several examples of these models can be found in the context of understanding
the behaviour of fusion plasmas ranging from the edge localized modes
(ELM) \cite{lebedev} to plasma turbulence \cite{ball}.

In spite of a great deal of experimental and theoretical research,
the sawteeth in tokamaks continue to be a subject of exploration.
For example, though numerical simulations \cite{aydemir} and some
experimental results \cite{snider} suggest total reconnection within
the safety factor $q=1$ surface during a sawtooth crash, there are
also experimental evidences \cite{campbell}, which indicate that
$q$ remains well below unity during a sawtooth cycle. Apparently,
there have been several important contributions to the understanding
of the sawtooth dynamics e.g. sawteeth with partial reconnection,
based on turbulent transport \cite{goedheer}. A Taylor relaxation
model of the sawteeth has also been considered by Gimblett and Hastie
\cite{gimblett}.

Here, we primarily focus on the dynamics of sawtooth oscillations
based on a low-dimensional dynamical model. We rigorously prove that
this dynamical model of the sawteeth based on thermal instability,
besides capturing the important physical aspects, does exhibit well
defined, isolated limit cycle oscillations, characteristics of self-excited
relaxation phenomena like the sawteeth. Alternatively, several authors
have proposed Hamiltonian models \cite{coppi2,coppi3,haas-1}, which
however can have infinite number of periodic solutions depending on
the starting points of the evolution with the same set of physical
parameters, which is rather inconsistent with the universal nature
of sawteeth for similar experimental conditions. These models, despite
having dissipation, possess a conserved quantity much like the Hamiltonian
of a conjugate system.

In Section II, we formulate the minimal dynamical model. In Section
III, we analyse the bifurcation and stability of the system, pointing
out the existence of an isolated and unique limit cycle and its global
stability. We complete this section by proving the uniqueness of the
limit cycle where we demonstrate the existence of an algebraic equation
for the limit cycle. Next, in Section IV, we address the issue of
relaxation oscillation and explore the parameter regime where the
limit cycle exhibits sawtooth-like oscillations. In Section V, with
the help of a renormalization group method, we prove that the invariant
manifold of the dynamical system is indeed the slow manifold of the
relaxation oscillation. In the Appendix, we outline the Hamiltonian
approach to this dynamical model.

\section{Dynamical modeling of sawtooth oscillations}

Typical sawtooth oscillations in small tokamaks $(R_{0}=1\,{\rm m})$
exhibit linear growth of central temperature with few milliseconds
of duration and rapid crash time of $\sim$ several microseconds,
in the Ohmic heating phase \cite{dubois,edwards,campbell}. Although
there are several other exotic cases viz. giant and monster sawteeth
\cite{campbell-2,campbell-3}, we shall limit our discussion to the
simpler type of sawteeth with a linear rise of the electron temperature.
The dynamical system, controlling the sawtooth oscillations of the
central electron temperature, then can be written as \cite{coppi1,coppi2,bora}\begin{eqnarray}
\frac{3}{2}n\frac{\partial T_{e}}{\partial t} & = & E_{\parallel}^{2}\sigma_{\parallel}-\nu_{L}(A,T_{e})nT_{e,}\\
\frac{\partial A}{\partial t} & = & \gamma(T_{e})A,\end{eqnarray}
 where $T_{e}$ is the central electron temperature expressed in energy
units, $A$ is the amplitude of the oscillation, $\nu_{L}(A,T_{e})$
is the rate of temperature redistribution, and $\gamma(T_{e})$ is
the growth rate of the relevant mode. The collisional parallel conductivity
$\sigma_{\parallel}\propto T_{e}^{3/2}$. In the above equations,
the particle density $n$ remains nearly constant during the sawtooth
cycle, which is consistent with experimental observations. We further
note that in the cases of sawtooth oscillations, we are going to consider,
the classical diffusion time \cite{hutchinson} for the plasma current
within the $q\leq1$ volume, $\tau_{J}=(r_{1}/d_{e})^{2}/\nu_{ei}$
(where $r_{1}$ is the radius of the $q=1$ surface, $d_{e}$ is the
plasma skin depth, and $\nu_{ei}$ is the electron-ion collision frequency),
is one order of magnitude higher than the sawtooth repetition time.
For example, in the Alcator C-Mod (MIT) machine, typically, $\tau_{J}\sim80$-$400\,{\textrm{msec}}$
for Ohmic regimes \cite{bombarda}, whereas the sawtooth period (crash
time being negligibly smaller than the period) $\tau_{{\textrm{st}}}\sim4\,{\textrm{msec}}$.
Therefore it can be safely assumed that the current redistribution
does not play any significant role and the corresponding $E_{\parallel}$
remains constant. We further assume that the pressure redistribution
parameter $\nu_{L}(A,T_{e})$ can be expressed with simple power laws
i.e. $\nu_{L}(A,T_{e})\propto T_{e}^{\alpha}A^{\sigma}$, where $\alpha$
and $\sigma$ are arbitrary constants.

With these considerations, we note that the second term in Eq.(1)
is responsible for the sawtooth crash. However, in absence of this
term the general solution of Eq.(1) is explosive. In particular, it
should include a loss term, which takes into account all other losses
e.g. radiation loss, electron thermal diffusion \cite{biskamp}, and
neoclassical loss \cite{hass-2} etc. With a diffusive loss term,
Eq.(1) can be rewritten as\begin{equation}
\frac{3}{2}n\frac{\partial T_{e}}{\partial t}=E_{\parallel}^{2}\sigma_{\parallel}-\nu_{L}^{0}nT_{e}^{\alpha}A^{\sigma}-\frac{3}{2}\frac{nT_{e}}{\tau_{E}},\end{equation}
 where $\tau_{E}$ is the overall plasma energy confinement time inside
the $q=1$ surface and $\nu_{L}^{0}$ is a proportionality constant.
In fact, the presence of the diffusive loss term can be understood
from the usual pressure gradient equation without taking into account
the instability which causes the ejection of plasma energy due to
sawtooth crashes,\begin{equation}
\frac{\partial p}{\partial t}=S_{{\rm Oh}}+\nabla\cdot\chi_{\perp}\nabla p,\end{equation}
 where $p\sim nT_{e}$ is the plasma pressure, $S_{{\rm Oh}}$ is
the Ohmic source term i.e. the heating power density per unit volume,
and $\chi_{\perp}$ is the plasma thermal conductivity perpendicular
to the magnetic field. Attributing all other losses to this diffusive
term, we can write \cite{itoh}\begin{equation}
(\nabla\cdot\chi_{\perp}\nabla)\sim-\tau_{E}^{-1}.\end{equation}
 Considering the fact that the plasma confinement time $\tau_{E}$
scales as $\sim a^{2}/\chi_{D}$, $a$ being the minor radius of the
torus and $\chi_{D}$, a diffusion coefficient which obeys Bohm-like
diffusion, i.e. $\chi_{D}=\rho_{i}\chi_{B}/a$, where $\chi_{B}=T_{e}/16eB$,
$\rho_{i}$ is the ion gyro-radius, and $B$ is the ambient toroidal
magnetic filed. Assuming that $T_{i}\approx T_{e}$, the last term
in Eq.(3) can be shown to scale as $\sim T_{e}^{5/2}$ and Eq.(3)
becomes\begin{equation}
\frac{3}{2}n\frac{\partial T_{e}}{\partial t}=E_{\parallel}^{2}\sigma_{\parallel}-\nu_{L}^{0}nT_{e}^{\alpha}A^{\sigma}-\beta nT_{e}^{5/2},\end{equation}
 where $\beta$ is a constant of proportionality. Note that the above
relation of electron temperature evolution closely resembles the corresponding
equation of sawtooth model by Kubota et. al. \cite{kubota}, based
on transport bifurcation.%
\begin{figure}
\begin{center}\includegraphics[%
  width=7cm,
  keepaspectratio]{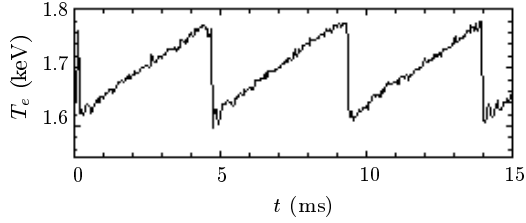}\end{center}

\caption{A typical sawtooth cycle in the Alcator C-Mod (MIT) machine (Shot
No. \#960130034 \cite{bombarda}).}
\end{figure}

We now consider Eq.(2) which controls the amplitude evolution of the
mode. During one sawtooth cycle of the profile shown in Fig.1, observed
in the Alcator C-Mod machine at MIT \cite{bombarda}. Since the sawtooth
crash time is much smaller than the sawtooth growth time, the mode
amplitude can be considered to have a `spike-like' behavior. In the
limiting case, the spike becomes a singularity and can be represented
by a $\delta$ function,\begin{equation}
A\sim\delta(t-t_{0}),\end{equation}
 centered around $t=t_{0}$. The corresponding relation for $T_{e}$
can be found out from Eq.(1) as\begin{equation}
T_{e}=t-\xi H(t-t_{0}),\end{equation}
 where $\xi$ is some constant and $H(t)$ is the Heaviside step function.
This behavior of $T_{e}$ shows a perfect sawtooth with a crash occurring
at $t=t_{0}$ and a discontinuity in $T_{e}$. In practice, the `spike-like'
behavior of amplitude $A$ can be approximated by an exponential function,\begin{equation}
A\sim\exp\bigl[-\mu(t-t_{0})^{2}\bigr],\end{equation}
 where $\mu$ is some large positive number. The corresponding $T_{e}$
is\begin{equation}
T_{e}\sim t-\xi_{1}\left\{ \sqrt{{2\mu}}(t-t_{0})\right\} +\xi_{2}.\end{equation}
 The related differential equation in $A$ is, therefore,\begin{equation}
\frac{dA}{dt}\simeq-2\mu(t-t_{0})A.\end{equation}
 The coefficient of $A$ on the right hand side of the above equation
can be viewed as $\gamma(T_{e}(t))$. So the complete set of dynamical
equations can now be written as \begin{eqnarray}
\frac{3}{2}n\frac{dT_{e}}{dt} & = & E_{\parallel}^{2}\sigma_{\parallel}-\nu_{L}^{0}nT_{e}^{\alpha}A^{\sigma}-\beta nT_{e}^{5/2},\\
\frac{dA}{dt} & = & \gamma_{0}\left(\frac{T_{e}}{T_{s}}-1\right)A,\end{eqnarray}
 where $T_{s}$ is the threshold temperature for the onset of the
relevant instability. We would like to emphasize that the model proposed
by Haas and Thyagaraja \cite{hass-2}, based on turbulent transport
essentially reduces to the above equations when the neoclassical losses
are discarded (they have a exclusive term to take care of the neoclassical
loss).

At this point, we would like to note that in reality, sawtooth oscillations
in magnetically confined plasmas are result of evolution of a `driven-dissipative'
system. During a sawtooth cycle, the plasma is heated by the current
through Ohmic dissipation which is essentially lost to the environment
after the plasma breakdown or the sawtooth crash. The sawtooth is
repeated as the plasma current continues to dissipate energy into
the plasma. Here, the plasma current is the driving mechanism of the
sawtooth oscillation through which, the energy is being continuously
dissipated. However, there is no driving source in Eq.(12) and the
oscillations represented by Eqs.(12-13) appears to be self-sustaining
i.e. a {}``self heating -- dissipating -- feedback -- heating''
cycle with a positive and negative damping mechanism through the terms
$\left(E_{\parallel}^{2}\sigma_{\parallel}-\beta nT_{e}^{5/2}\right)$.
The justification for Eq.(12) comes from the fact that we are looking
only at the power balance equation for the confined plasma \cite{stacey},
not at the full burn-cycle of the plasma confinement. So, as far as
the plasma power balance is concerned, we assume that there is a perpetual
source of energy in the form of the plasma current provided externally
through the loop voltage, which is the driving mechanism and an infinite
sink which drains away the energy dissipated through the sawtooth
crash. As the whole cycle repeats, the power balance equation becomes
self-sustaining.

We further note that during the sawtooth ramp phase, there is a definite
change in the magnetic configuration, which rearranges itself after
the sawtooth crash either through a partial or complete reconnection
\cite{biskamp}. As this magnetic reorientation causes the electron
temperature to drop during the sawtooth crash, we have dynamically
modeled this event by the second term $\left(\nu_{L}^{0}nT_{e}^{\alpha}A^{\sigma}\right)$,
in Eq.(12).

\section{Stability and bifurcation \emph{--- limit cycle oscillation}}

In this section, we examine the possibility of existence of periodic
orbits of Eqs.(12) and (13), in particular existence of any limit
set. This is to be viewed in contrast to the Hamiltonian formalism
(see Appendix), which always admits periodic solutions, though the
solutions are not unique.

We begin by expressing these dynamical equations in a generic dimensionless
form. The electron temperature $T_{e}$ is normalized by the threshold
temperature $T_{s}$ and define $x=T_{e}/T_{s}$. We further define
the other dimensionless quantities as\begin{equation}
\frac{2}{3}\frac{\nu_{L}^{0}T_{s}^{\alpha-1}}{S_{{\rm Oh}}}A^{\sigma}=y,\qquad\frac{2}{3}\frac{\beta}{S_{{\rm Oh}}}T_{s}^{3/2}=a.\end{equation}
 The quantity $y$ denotes the normalized amplitude of the mode and
the quantity $S_{{\rm Oh}}$ is the Ohmic source term corresponding
to the threshold temperature $T_{s}$,\begin{equation}
S_{{\rm Oh}}=\frac{2}{3}\frac{E_{\parallel}^{2}\sigma_{s\parallel}}{nT_{s}},\end{equation}
 the quantity $\sigma_{s\parallel}$ being the collisional conductivity
at $T_{e}=T_{s}$. With these definitions, the set of Eqs.(12) and
(13) can be written in a generic form as \begin{eqnarray}
\dot{x} & = & x^{(3/2)}(1-ax)-x^{\alpha}y,\\
\dot{y} & = & \rho(x-1)y,\end{eqnarray}
 where $\rho=\sigma\gamma_{0}/S_{{\rm Oh}}$. Note that in writing
these equations we have re-scaled the time as $t\rightarrow S_{{\rm Oh}}t$.
As we shall see shortly that the constant $a<1$ while $\rho$ which
is related to the growth rate can be quite large i.e. $\gg1$. The
dots represent the derivatives with respect to the rescaled time $t$.
\begin{figure}
\begin{center}\includegraphics[%
  width=7cm,
  keepaspectratio]{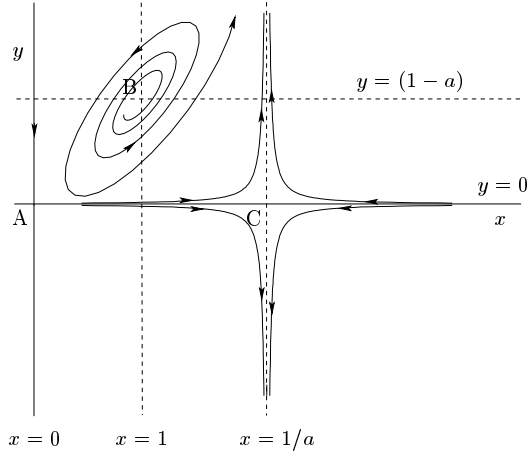}\end{center}

\caption{Phase portrait of Eqs.(16) and (17). The arrows denote the direction
of flow.}
\end{figure}

In order to examine the bifurcation diagram of Eqs.(16) and (17) and
existence of limit cycle oscillation, it is instructive to construct
a phase portrait of the equations, which is shown in Fig.2. The arrows
on the curves show the direction of time. As can be seen, there are
three fixed points, `A', `B', and `C' located at $(x,y)=(0,0)$, $(1,1-a)$,
and $(1/a,0)$, respectively. Point `A' at the origin is a non-isolated
fixed point. Point `C' is a saddle point. The only fixed point of
interest is the point `B' which is either a stable or unstable point
depending on values of $\alpha$ and $a$. We are also interested
in this fixed point as this is the only point which is in the positive
quadrant as the variables $x$ and $y$ may take only positive values.
In what follows, we shall show that surrounding the fixed point `B',
we can construct a positively invariant region \cite{hale} lying
entirely in the first quadrant.%
\begin{figure}
\begin{center}\includegraphics{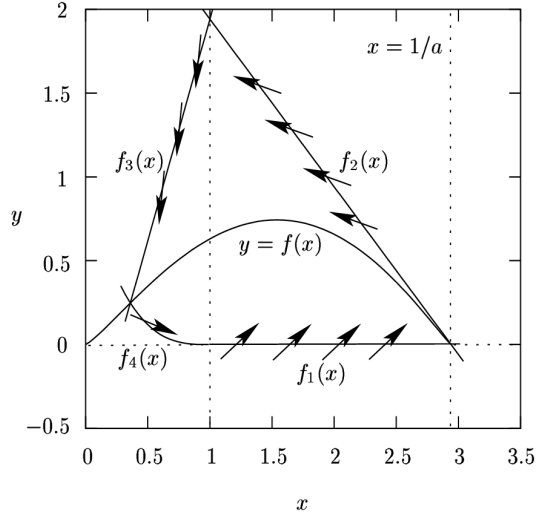}\end{center}

\caption{A positively invariant region for Eqs.(16) and (17).}
\end{figure}

\subsection{Limit cycle oscillation}

Consider now, a set of curves $f_{i}(x)$ given by the equations\begin{eqnarray}
f_{1}(x) & = & 0,\\
f_{2}(x) & = & \mathcal{C}\left(\frac{1}{a}-x\right),\\
f_{3}(x) & = & x\left[\mathcal{C}\left(\frac{1}{a}-1\right)+\delta\right]-\delta,\\
f_{4}(x) & = & \mathcal{D}(1-x)^{2\mu+1},\end{eqnarray}
 where ${\mathcal{\mathcal{C}}}$ can be a large positive number,
$\delta$ is a small positive number, and $\mathcal{D}$ and $\mu$
are positive arbitrary constants. The nullclines of the Eqs.(16) and
(17) are given by the equations\begin{eqnarray}
x & = & 1,\\
y & = & x^{3/2-\alpha}(1-ax)=f(x).\end{eqnarray}
 If we now span a region (see Fig.3) described by the set of curves
$f_{i}(x)$s through the points $(1,0)$, $(1/a,0)$, $(1,\mathcal{C}(1/a-1))$,
and $(x',y')$ --- the points of intersection of the curves $f_{i}(x)$
and $f(x)$, $(x',y')$ being the common point of intersection of
$f_{3,4}(x)$ and $y=f(x)$, it can be shown by comparing the slopes
of the curves $f_{i}(x)$s with that of the orbit described by Eqs.(16)
and (17), that the flow of the orbit of Eqs.(16) and (17) always crosses
\emph{into} the closed region, anticlockwise \cite{hale}. Thus the
flow of the orbit of Eqs(16) and (17) is always bounded and the closed
region in Fig.3 constitutes a positively invariant region for Eqs.(16)
and (17). An example set of parameters for this positively invariant
region are $a=0.34$, $\alpha=1/2$, ${\mathcal{\mathcal{C}}}=1$,
$\mu=1$, $\mathcal{D}=0.923$, and $\delta=1/2$, the point $(x',y')=(0.322,0.287)$.

What needs to shown now is the fact that for the permitted parameter
regime, the fixed point `B' is an unstable point. We linearize the
set of equations (16) and (17) around the fixed point `B' $=(1,1-a)$
and the eigenvalues of the linear part of these equations at $(1,1-a)$
are\begin{equation}
\lambda_{1,2}=\frac{1}{2}\left(\tau\pm\sqrt{{\tau^{2}-4\Delta}}\right),\end{equation}
where\begin{eqnarray}
\tau & = & \frac{3}{2}-\frac{5}{2}a-\alpha(1-a),\\
\Delta & = & \rho(1-a),\end{eqnarray}
 and the stability of the equilibrium point `B' depends on the values
of $\tau$ and $\Delta$. As $\rho\gg1$ and $a<1$, we can find that
the point `B' is an unstable spiral, if\begin{equation}
a<\left(\frac{3/2-\alpha}{5/2-\alpha}\right)=a_{c},\:\alpha<\frac{3}{2}=\alpha_{c}.\end{equation}
 The existence of the limit cycle is now evident according to the
Poincar\'{e}-Bendixson condition \cite{strogatz}.

The point `B' also becomes unstable for $\alpha>5/2$ provided $a>1$,
however the latter is excluded by our physical conditions. As the
growth rate of the instability should be large enough to cause a sawtooth
crash, $\rho\gg1$. This also ensures that the fixed point `B' is
an unstable spiral. Also, we observe that as $a\rightarrow1$, $\alpha\rightarrow\infty$.
In fact, at $a=a_{c}$ (for a certain $\alpha<\alpha_{c}$), a supercritical
Hopf bifurcation occurs \cite{strogatz,hale}. The bifurcation surface
for Eqs.(16) and (17) is shown in Fig.4. Note the steepening of the
surface $f(x,\alpha)$, given by Eq.(23), as $\alpha\rightarrow3/2$
and the disappearance of the fixed point `A'$=(0,0)$ at the bifurcation
point.%
\begin{figure}
\begin{center}\includegraphics{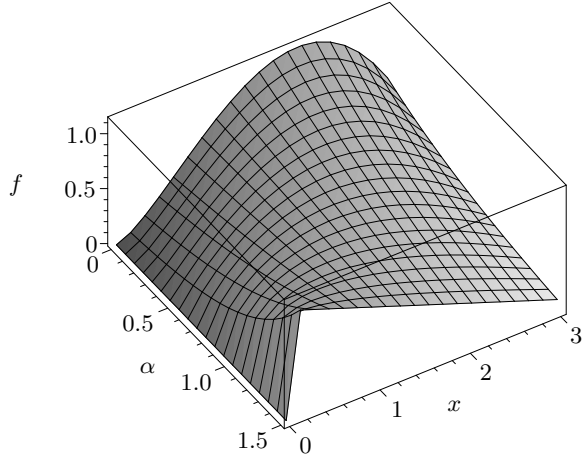}\end{center}

\caption{Bifurcation diagram for the set of equations (16) and (17).}
\end{figure}
 In Fig.5, we show the limit cycle for a set of parameters $\alpha=1$,
$a=0.33$, and $\rho=1$. At the birth of the limit cycle after the
supercritical Hopf bifurcation, the period of the limit cycle is given
by the Hopf period,\begin{equation}
\tau_{{\rm Hopf}}=\frac{4\pi}{\sqrt{{4\Delta-\tau^{2}}}}\approx\frac{2\pi}{\sqrt{{\rho(1-a)}}}.\end{equation}
\begin{figure}
\begin{center}\includegraphics{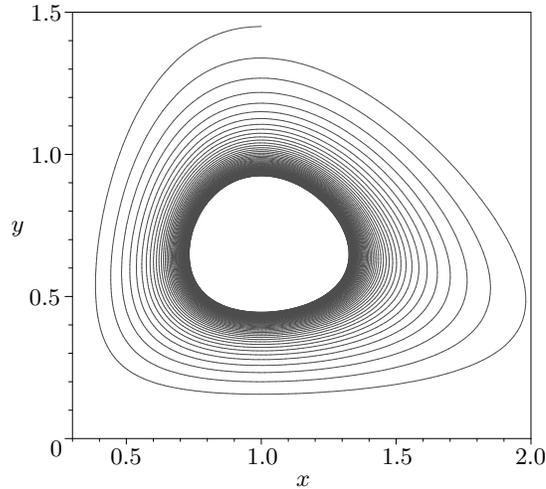}\end{center}

\caption{The limit cycle for Eqs.(16) and (17) subject to condition (27).}
\end{figure}

\subsection{Local and global stability of the limit cycle}

Though the phenomena of sawtooth oscillations in magnetically confined
plas\-mas suggests existence of a `driven-dissipation' type limit
cycle oscillation, experimentally observed sawtooth oscillations are,
in general, perturbed by local irregularities and fluctuations. As
limit cycles are usually characterised by a flow field which converges
to the cycle everywhere, it is important to classify the \emph{orbital
stability} of the limit cycle in order to emphasize its response to
noise.

The classical method for studying orbital stability of limit cycles
involves eigenvalues of the Poincar\'e map \cite{jackson}. Several
other variants also exist in literature, for studying local stability
\cite{tomita,kurrer}, among which the most common one is the theory
of Floquet multipliers or the \emph{monodromy matrix} method, where
one follows the short-time evolution of a small perturbation $\delta\bm x$
along the limit cycle. Another widely applicable method is associated
with the direction of \emph{slowest decay} \cite{wolf}. Here, we
employ a method based on applying a small perturbation $\delta\bm x(t)$
to the non-stationary reference solution i.e. limit cycle of the flow
$\bm x_{0}(t)$ and following the directions tangential and transverse
to the flow along the perturbed trajectory $\bm x(t)=\bm x_{0}(t)+\delta\bm x(t)$
\cite{ali}. 

Consider the dynamical system \begin{equation}
\frac{d}{dt}\bm x(t)=\bm F[\bm x(t)].\end{equation}
If the perturbations $\delta\bm x(t)$ are small enough, the evolution
can be linearised \begin{equation}
\frac{d}{dt}\delta\bm x=\bm A\,\delta\bm x,\qquad A_{ij}=\left(\frac{\partial\bm F_{i}}{\partial x_{j}}\right)_{\bm x_{0}(t)}.\end{equation}
The linearised system (30) is transformed from the fixed basis $\bm x$
to the rotating, orthogonal basis $\bm x'$ through a unitary rotation
matrix $\bm U$, $\delta\bm x'=\bm U\,\delta\bm x,\delta\bm x=\bm U^{-1}\,\delta\bm x'$,
so that the linearised system (30) becomes\begin{equation}
\frac{d}{dt}\delta\bm x'=\bm B\,\delta\bm x',\end{equation}
where\begin{equation}
\bm B=\bm U\bm A\bm U^{-1}+\frac{d\bm U}{dt}\bm U^{-1}.\end{equation}
is the stability matrix.

For a two-variable system like ours {[}Eqs.(16-17){]},\begin{eqnarray}
\dot{x} & = & f(x,y),\\
\dot{y} & = & g(x,y),\end{eqnarray}
the stability matrix $\bm B$ is given by\begin{equation}
\bm B=\left(\begin{array}{cc}
\lambda_{tt} & \lambda_{tn}\\
\lambda_{nt} & \lambda_{nn}\end{array}\right),\end{equation}
where\begin{eqnarray}
\lambda_{tt} & = & [f^{2}f_{x}+g^{2}g_{y}+fg(f_{y}+g_{x})]/(f^{2}+g^{2}),\\
\lambda_{tn} & = & [(g^{2}-f^{2})(f_{y}+g_{x})+2fg(f_{x}-g_{y})]/(f^{2}+g^{2}),\\
\lambda_{nt} & = & 0,\\
\lambda_{nn} & = & [f^{2}g_{y}+g^{2}f_{x}-fg(f_{y}+g_{x})]/(f^{2}+g^{2}).\end{eqnarray}
 are the rates of convergence of the flow from the perturbed trajectory
into the limit cycle. Parameter $\lambda_{nn}$ represents the rate
of convergence normal to the flow and $\lambda_{tt}$ represents convergence
tangential to the flow, which essentially averages to zero over a
full cycle. The tangential and normal perturbations are only partially
coupled as $\lambda_{nt}=0$, which means that a tangential perturbation
remains tangential whereas a normal perturbation couples to the tangential
perturbations (nonzero $\lambda_{tn}$). So a perturbation is unstable
or stable depending on whether $\lambda_{ij}$ is positive or negative.
The global stability measures can be found out from the Lyapunov exponent,
which can be obtained by phase averaging the $\lambda_{ij}$s over
an orbit with period $T$ \cite{ali},\begin{equation}
\Lambda_{ij}=\frac{1}{T}\oint\lambda_{ij}(\phi)\, d\phi,\end{equation}
corresponding to a particular perturbation. A negative $\Lambda$
would imply global stability. In practice we use a discretised version
of Eq.(40),\begin{equation}
\Lambda_{ij}=\frac{1}{m}\sum_{k=1}^{m}\lambda_{ij}(t_{0}+k\,\Delta t),\qquad m\,\Delta t=T.\end{equation}

In Fig.6, we show the stability properties of the limit cycle oscillation
of Eqs.(16-17) for a set of parameters $\alpha=1,a=0.33$, and $\rho=100$,
which produces a relaxation type oscillation (see Fig.8). As one can
see that the limit cycle consists of unstable and stable parts corresponding
to the normal perturbations. The limit cycle is globally stable as
indicated by the global stability parameter, the Lyapunov exponent
$\Lambda_{nn}=-0.00431$. The plot of $\Lambda_{tt}$ shows the tangential
stability which is essentially due to acceleration and deceleration
of the flow showing a relaxation mechanism. As expected, the average
$\Lambda_{tt}\approx0$ within the computational accuracy. The coupling
matrix element $\lambda_{tn}$ represents the effect of perpendicular
perturbation on the oscillator phase. During $\lambda_{tn}>0$, the
perpendicular perturbation leads to advancement of the phase of the
oscillator and $\lambda_{tn}<0$ indicates a phase delay.%
\begin{figure}
\begin{center}\includegraphics[%
  width=0.5\textwidth]{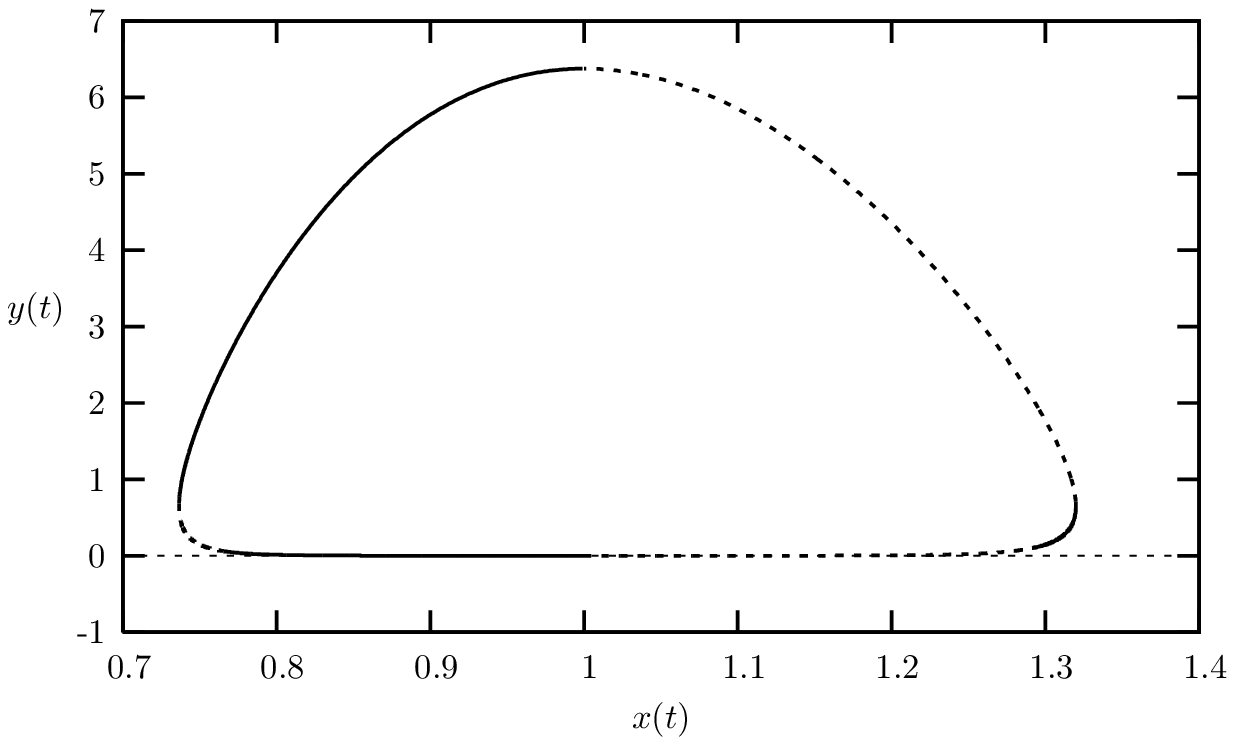}\hfill{}\includegraphics[%
  width=0.5\textwidth]{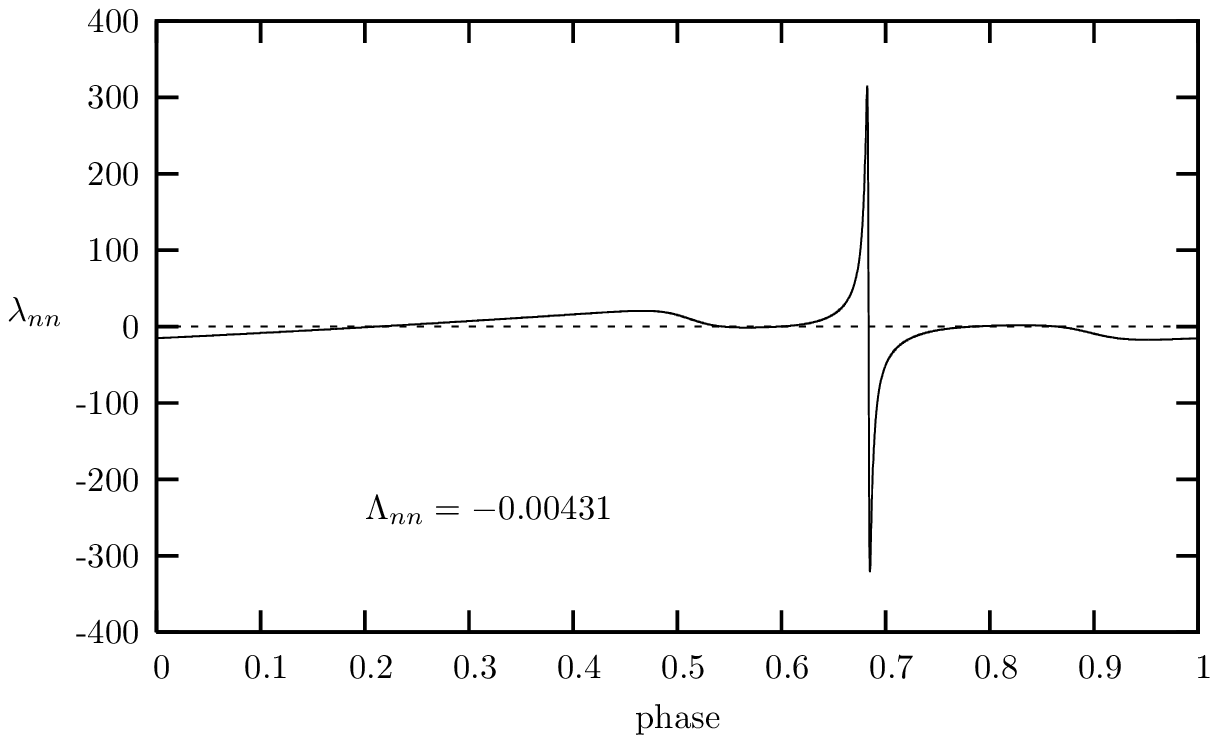}\end{center}

\begin{center}\includegraphics[%
  width=0.5\textwidth]{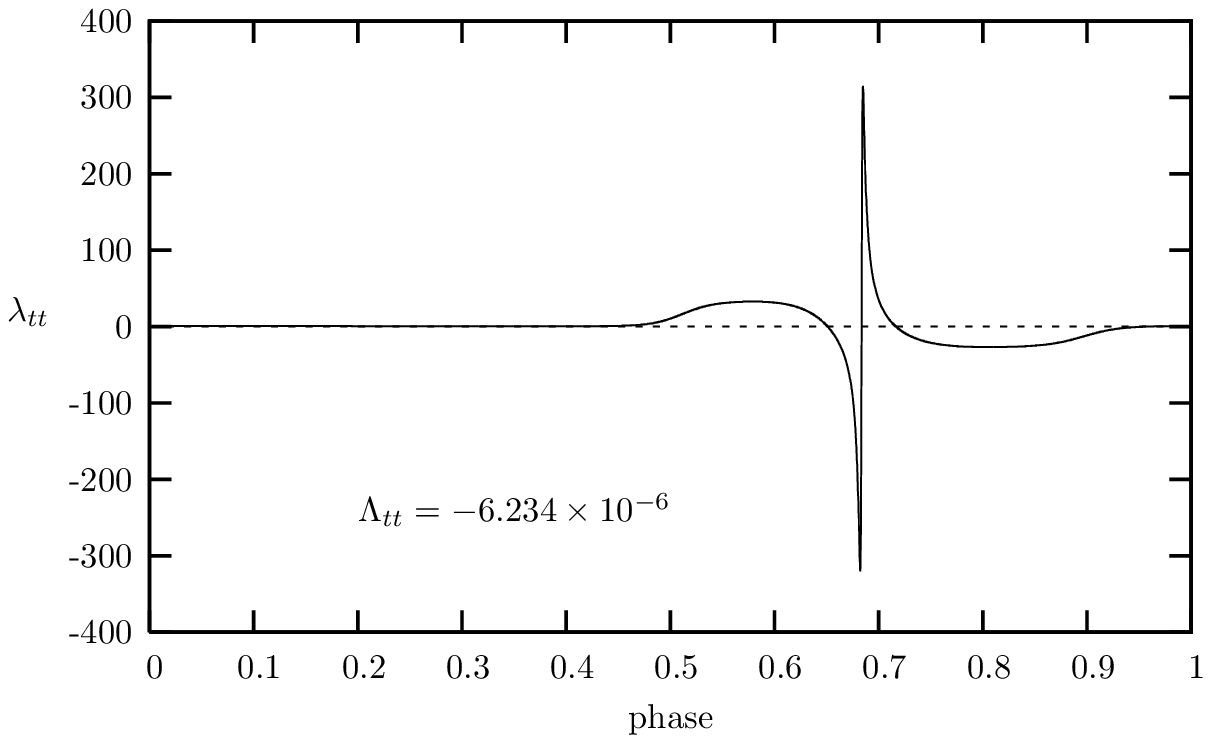}\hfill{}\includegraphics[%
  width=0.5\textwidth]{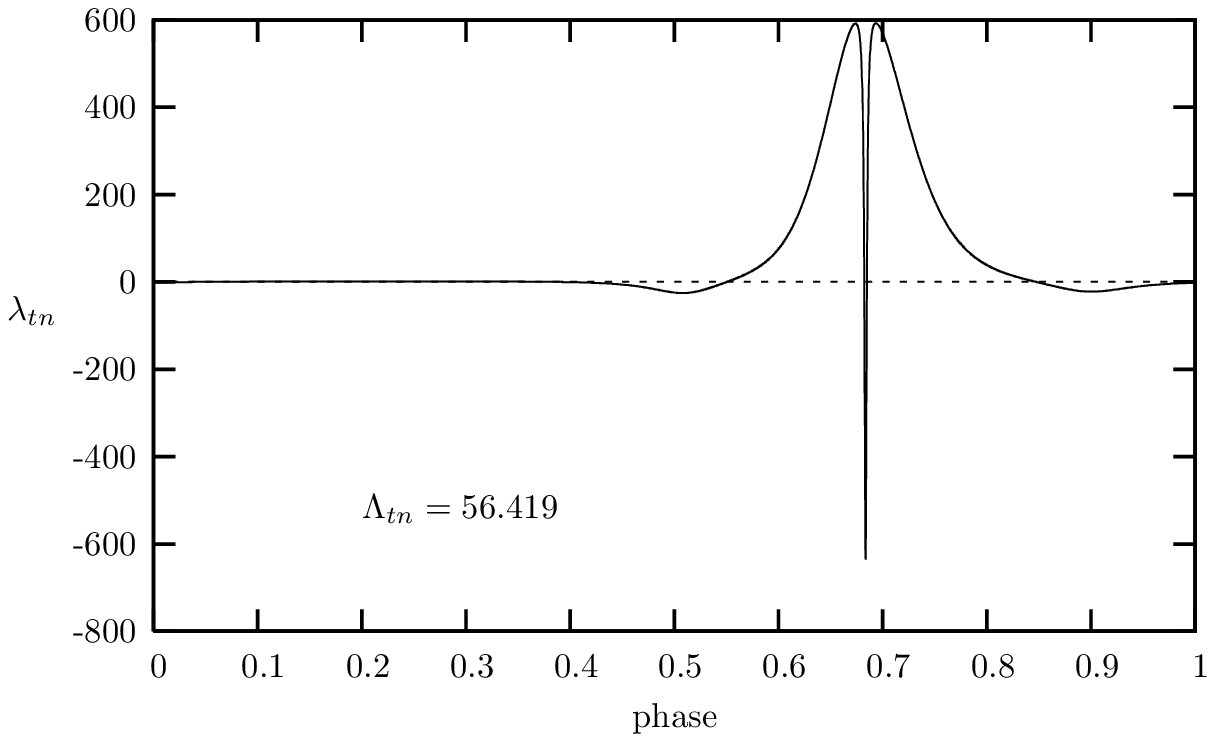}\end{center}

\caption{Local and global stability of the limit cycle of Eqs.(16-17) away
from the Hopf bifurcation point. Clockwise from top, in the first
plot, the unstable (dotted) and stable (solid) parts of the limit
cycle in the phase plane are shown. The variations of $\lambda_{nn},\lambda_{tn}$,
and $\lambda_{tt}$ are shown over one phase of the limit cycle in
the second, third, and fourth plots.}
\end{figure}

\subsection{Equation of the limit cycle}

In this subsection, we find out an equation for the limit cycle enclosing
the critical point `B' in terms of a series expansion. Here, we employ
a method due to Giacomini and Viano \cite{giacomini}, which is based
on the fact that if we consider the partial differential equation\begin{equation}
P\frac{\partial V}{\partial x}+Q\frac{\partial V}{\partial y}=\left(\frac{\partial P}{\partial x}+\frac{\partial Q}{\partial x}\right)V,\end{equation}
 corresponding to the two-dimensional dynamical system\begin{equation}
\dot{x}=P(x,y),\qquad\dot{y}=Q(x,y),\end{equation}
 there exists a unique convergent power series solution of Eq.(42)
in some region $\mathfrak{R}$ containing a non-degenerate critical
point $P_{0}$ of node or focus (spiral) type and that this solution
satisfies\begin{equation}
V(x,y)=0\end{equation}
 on any limit cycle contained in $\mathfrak{R}$ which encloses $P_{0}$.
In our case, the equations (16) and (17) have an isolated limit cycle
enclosing the critical point located at $(1,1-a)$. We have already
seen that subject to the conditions (27), the equilibrium point `B'
is an unstable spiral. Therefore, there exists a unique series solution
of Eq.(42).

As the cases, where an explicit series solution can be calculated
analytically, are limited to polynomial forms for the functions $P(x,y)$
and $Q(x,y)$, we write the Eqs.(16,17) with a transformation of variable
$x\rightarrow x^{2}$, so that our system of dynamical equations become\begin{eqnarray}
\dot{x} & = & \frac{1}{2}x^{2}(1-ax^{2})-\frac{1}{2}x^{2\alpha-1}y,\\
\dot{y} & = & \rho(x^{2}-1)y.\end{eqnarray}
 Without loss of any generality, the constant $\alpha$ can be set
to unity, for which, the right hand sides of Eq.(45,46) become polynomials
in $x$ and $y$. Following Giacomini and Viano \cite{giacomini},
we now look for a power series\begin{equation}
V(x,y)=\sum_{n=0}^{\infty}v_{n}(x,y),\end{equation}
 where $v_{n}(x,y)$ is a homogeneous polynomial of degree $n$,\begin{equation}
v_{n}(x,y)=\sum_{k=0}^{n}c_{n-k,k}x^{n-k}y^{k}.\end{equation}
 In practice, we work on the truncated sum\begin{equation}
V(x,y)=\sum_{n=0}^{N}v_{n}(x,y),\end{equation}
 at order $N$ and try to solve for the coefficients $c_{n-k,k}$
from set of simultaneous equations generated from partial derivatives
of Eq.(42) evaluated at the critical point $P_{0}=(1,1-a)$. The equation
of the limit cycle is thus given by%
\begin{table}

\caption{The coefficients $c_{n-k,k}$ for order $N=8$.}

\begin{center}\begin{tabular}{|crr|rcrr|rcrr|rcr|}
\hline 
$c_{8,0}$&
$1.0$&
&
&
$c_{4,1}$&
$84.977$&
&
&
$c_{5,3}$&
$-2.945$&
&
&
$c_{3,5}$&
$0.037$\tabularnewline
$c_{7,0}$&
$-6.724$&
&
&
$c_{3,1}$&
$-89.203$&
&
&
$c_{4,3}$&
$17.662$&
&
&
$c_{2,5}$&
$-0.494$\tabularnewline
$c_{6,0}$&
$19.626$&
&
&
$c_{2,1}$&
$56.673$&
&
&
$c_{3,3}$&
$-50.909$&
&
&
$c_{1,5}$&
$-5.958$\tabularnewline
$c_{5,0}$&
$-32.122$&
&
&
$c_{1,1}$&
$-16.609$&
&
&
$c_{2,3}$&
$60.445$&
&
&
$c_{0,5}$&
$-11.421$\tabularnewline
$c_{4,0}$&
$31.508$&
&
&
$c_{6,2}$&
$-1.691$&
&
&
$c_{1,3}$&
$-28.992$&
&
&
$c_{2,6}$&
$0.003$\tabularnewline
$c_{3,0}$&
$-17.578$&
&
&
$c_{5,2}$&
$12.896$&
&
&
$c_{0,3}$&
$-5.502$&
&
&
$c_{1,6}$&
$3.010$\tabularnewline
$c_{2,0}$&
$4.296$&
&
&
$c_{4,2}$&
$-43.927$&
&
&
$c_{4,4}$&
$-1.795$&
&
&
$c_{0,6}$&
$8.242$\tabularnewline
$c_{1,0}$&
$0.0$&
&
&
$c_{3,2}$&
$90.659$&
&
&
$c_{3,4}$&
$10.090$&
&
&
$c_{1,7}$&
$-0.550$\tabularnewline
$c_{7,1}$&
$-2.723$&
&
&
$c_{2,2}$&
$-96.759$&
&
&
$c_{2,4}$&
$-13.801$&
&
&
$c_{0,7}$&
$-3.522$\tabularnewline
$c_{6,1}$&
$17.608$&
&
&
$c_{1,2}$&
$44.352$&
&
&
$c_{1,4}$&
$12.472$&
&
&
$c_{0,8}$&
$0.658$\tabularnewline
$c_{5,1}$&
$-51.146$&
&
&
$c_{0,2}$&
$-2.460$&
&
&
$c_{0,4}$&
$10.162$&
&
&
&
\tabularnewline
\hline
\end{tabular}\end{center}
\end{table}
\begin{equation}
V(P_{0})=0.\end{equation}

For a set of parameters $\alpha=1$, $a=0.33$, and $\rho=1$, satisfying
conditions (27), we calculate the coefficients $c_{n-k,k}$ with the
help of the computer algebra system of Waterloo Maple \cite{maple}
and find that a close curve fairly coinciding the numerically calculated
limit cycle exists for an order as low as $N=8$. The corresponding
curves of the the limit cycle are shown in Fig.7. Note that the value
of $\rho=1$ does not necessarily produce a sawtooth, we however have
chosen this value so as to find a close curve for $V(x,y)$ at the
lowest possible order. The corresponding coefficients $c_{n-k,k}$
are tabulated in Table.1.%
\begin{figure}
\begin{center}\includegraphics[%
  width=7cm,
  keepaspectratio]{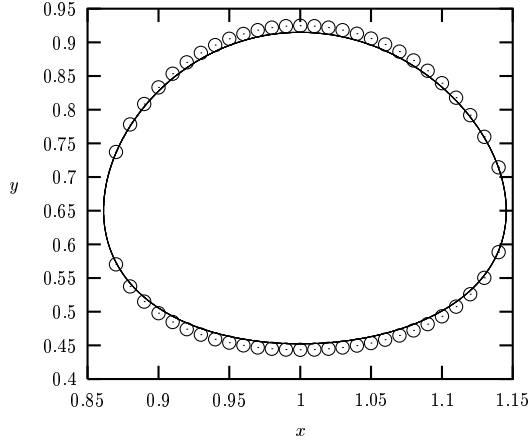}\end{center}

\caption{The limit cycle for Eqs.(32) and (33) as determined by the Eq.(37)
(open circles) and numerically calculated one (solid).}
\end{figure}

For the sake of completeness, we review the equivalent Hamiltonian
approximation of Eqs.(16) and (17), in the Appendix.

\section{Relaxation oscillation and sawtooth disruptions}

Relaxation oscillations are characterized by two very different time
scales. As we can see from the phase portrait of Eqs.(16,17) in Fig.2,
relaxation oscillation, if any, must be away from the Hopf bifurcation
point i.e. $a=a_{c}$. So, we must push the system toward the boundary
of the invariant region, defined by the equilibrium points `A' and
`C', around the bifurcation point $(1,1-a)$. At the same time, it
must be noted that far away from the bifurcation point i.e. for $a\ll a_{c}$,
the orbit of the limit cycle will be \emph{swept} away in the flow
around the saddle point `C'.

A sawtooth trigger is caused by a sudden increase in a \emph{fast}
variable ($y$), as have already been pointed out in Sec.II, which
remains almost close to zero for the rest of the sawtooth cycle. This
ensures an almost linear rise of the sawtooth i.e. in the \emph{slow}
variable, the variable $x$, in our case. As shown in Fig.8, Eqs.(16,17)
do indeed exhibit relaxation oscillations of sawtooth-type for an
appropriate set of parameters. %
\begin{figure}
\begin{center}\includegraphics{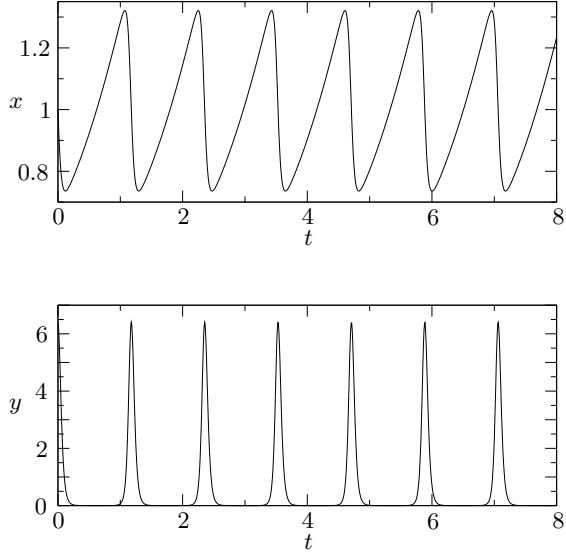}\end{center}

\caption{Limit cycle, relaxation oscillation (sawtooth) of Eqs.(16) and (17).
The parameters in this particular case are $\alpha=1$, $a=0.33$,
and $\rho=100$.}
\end{figure}
 The phase diagram of the corresponding limit cycle is shown Fig.8.
\begin{figure}
\begin{center}\includegraphics{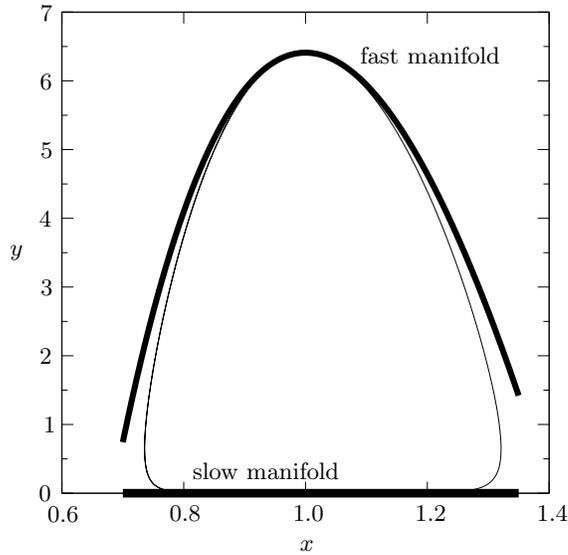}\end{center}

\caption{The phase diagram of the limit cycle (the thin closed curve) corresponding
to the sawtooth oscillations shown in Fig.8. The fast and slow manifolds
are shown in thick lines.}
\end{figure}
 As expected the fast variable $y$ remains nearly zero for most of
the time except for a vary short time, when it attains very high value,
which causes the sawtooth crash.

The physical mechanism of this relaxation oscillation can be understood
in terms of similar oscillations in model sandpile \cite{kadanoff}.
The corresponding variable in case of a sandpile is $\phi(x)$, the
height of the pile at a position $x$. When sand is continuously added
at the top of the pile, an instability occurs when the local slope
of the pile becomes too large ($\nabla\phi$ greater than a critical
value), causing an \emph{avalanche}, which consists of rearrangements
on the surface of the pile. In case of sawtooth oscillation, energy
from the source flows at the rate $\sim x^{3/2-\alpha}$ {[}see Eq.(16-17){]},
causing the temperature to rise steadily, which then triggers an `avalanche'
after the temperature rises above a critical value.

\subsection{The slow and fast manifolds}

On the limit cycle, if we start from a point $x<1$ on the slow manifold
(see Fig.9), the damping term in Eq.(16) i.e. $(x^{\alpha}y)$ remains
very small as $y\approx0$. As the constant $a<1$, $dx/dt$ remains
positive and the slow variable $x$ increases in time. At the same
time $dy/dt<0$ and $y$ further decreases as long as $x<1$. However
as $x$ increases beyond unity, $dy/dt$ becomes positive and $y$
increases as $\sim\exp(t^{2})$, which causes the damping term in
Eq.(16) to dominate over the growth term, triggering a crash. As soon
as $x$ again falls below unity, $y$ rapidly drops {[}see Eq.(17){]}
and the whole cycle repeats. We expect that in the parameter region
where this relaxation oscillations occur, the sawtooth growth time
($\tau_{{\rm st}}$) should be independent of the growth rate of the
instability $\rho\gg1$, whereas, the sawtooth crash time ($\tau_{{\rm c}}$)
should be $\propto\rho^{-1}$.

To examine further, we note that on the slow manifold we can change
the scale by introducing a small quantity $\epsilon$ which defines
the fast time scale $t'=t/\epsilon$. As the growth rate $\rho$ can
be large and $y\approx0$ on the slow manifold, rescaling them as\begin{equation}
y=\epsilon\eta,\qquad\rho=\delta/\epsilon,\ \end{equation}
we can write Eqs.(16,17) as\begin{eqnarray}
\dot{x} & = & x^{3/2}(1-ax)-\epsilon x^{\alpha}\eta,\\
\epsilon\dot{y} & = & \delta(x-1)\eta.\end{eqnarray}
Taking the limit $\epsilon\rightarrow0$, we obtain the slow manifold
as\begin{eqnarray}
y & = & 0,\\
\dot{x} & = & x^{3/2}(1-ax),\end{eqnarray}
so that the sawtooth period is essentially given by the time that
the variable spends on the slow manifold,\begin{equation}
\tau_{{\rm st}}\simeq\int_{x_{1}}^{x_{2}}\frac{dx}{x^{3/2}(1-ax)},\end{equation}
where $x_{1,2}$ are the points of intersection of the limit cycle
with the nullcline $y=f(x)$ {[}see Eq.(23){]}. As the drop in electron
temperature i.e. $\Delta T_{e}\ll1$ during a sawtooth crash, so is
the drop in the corresponding dynamical variable $\Delta x$ and hence
the amplitude of the sawtooth oscillation. Expanding Eq.(56) around
the mean value of $x=1$, we can further estimate the sawtooth period
in terms of its amplitude,\begin{equation}
\tau_{{\rm st}}\sim(1-a)\Delta x.\end{equation}

On the fast manifold, as $y$ can attain large values, scaling $y$
as $\eta/\epsilon$ and time $t$ as $t'\epsilon$, we have from Eqs.(16,17),\begin{eqnarray}
x' & = & \epsilon x^{3/2}(1-ax)-x^{\alpha}\eta,\\
\eta' & = & \delta(x-1)\eta,\end{eqnarray}
where the $(')$ denotes derivatives with respect to the fast time
scale $t'$. We obtain the fast manifold by letting $\epsilon\rightarrow0$,\begin{eqnarray}
\dot{x} & = & -x^{\alpha}y,\\
\dot{y} & = & \rho(x-1)y,\end{eqnarray}
which is essentially given by\begin{equation}
y=\rho(\ln x-x+1)+y_{m},\end{equation}
for $\alpha=1$, $y_{m}$ being the maximum value of $y$ on the fast
manifold which occurs at $x=1$. Both these manifolds are shown in
Fig.9. The sawtooth crash time $\tau_{{\rm c}}$ can be approximately
estimated from Eq.(59) and (61),\begin{equation}
\tau_{{\rm c}}\simeq\int_{x_{2}}^{x_{1}}\frac{dx}{x[\rho(\ln x-x+1)+y_{m}]}.\end{equation}
As expected, in the parameter space of relaxation oscillation ($\rho\gg1$),
the sawtooth period is essentially independent of the growth rate
$\rho$ while the crash time is inversely proportionate to it (see
Fig.10). Note that with the index $p=0.5$, the total repetition time
of the sawtooth $\tau=\tau_{{\rm st}}+\tau_{{\rm c}}$ closely follows
the Hopf period as given by Eq.(28) when the oscillations remain purely
sinusoidal for $\rho\sim1$. %
\begin{figure}
\begin{center}\includegraphics{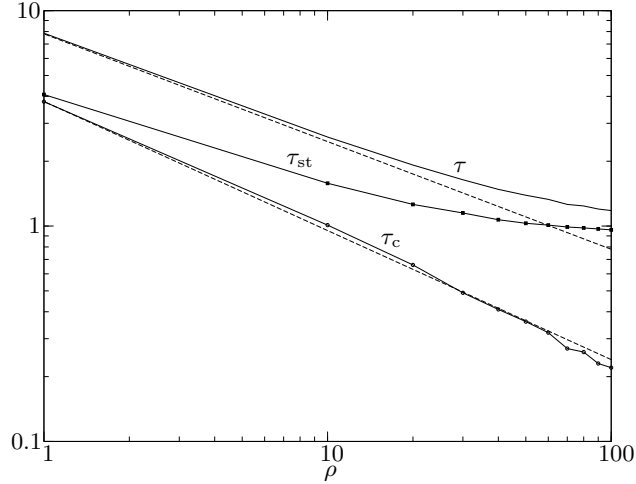}\end{center}

\caption{As $\rho$ becomes $\gg1$, the sawtooth period $\tau_{{\rm st}}$
becomes independent of $\rho$. The crash time $\tau_{{\rm c}}$ however
remains inversely proportionate to $\rho$, $\tau_{{\rm c}}\propto\rho^{-p}$.
The dotted lines are the fitted curves for $\tau_{{\rm c}}$ and $\tau=\tau_{{\rm st}}+\tau_{{\rm c}}$
with the index $p=0.6$ and $0.5$, respectively.}
\end{figure}

\subsection{Simulations of sawteeth}

In Fig.11, we show the reproduction two sets of sawtooth oscillations
from the Alcator C-Mod tokamak at MIT \cite{bombarda}, which has
an effective major and minor radius of $0.68\,{\rm m}$ and $0.22\,{\rm m}$.
For the two experimental shots, (\emph{a}) Shot No. \#960130034 and
(\emph{b}) \#960127012, the threshold temperatures $T_{s}$ are taken
to be the average temperatures of the sawteeth, which are, $1.7$
and $1.4\,{\rm keV}$. The sawtooth time periods are respectively
4.63 and 3.6 milliseconds. The constant applied electric field is
given by $E_{\parallel}=0.291$ and $0.352\,{\rm volt/m}$ which correspond
to the Ohmic source term $S_{{\rm Oh}}\simeq83.08$ and $62.09\,{\rm s^{-1}}$
for the value of Coulomb logarithm $\ln\Lambda\simeq17$. The parallel
plasma conductivity corresponding to the threshold temperature $T_{s}$
is given by \begin{equation}
\sigma_{s\parallel}=9.71\times10^{3}Z_{{\rm eff}}\,\ln\Lambda\, T_{s}^{3/2}.\,{\rm si/m}\end{equation}
The temperature drop $\Delta T_{e}$ during the sawtooth crash are
$0.18$ and $0.14\,{\rm KeV}$, for these two cases. The line-averaged
plasma density in the Alcator C-Mod is between $1\times10^{20}$ and
$3\times10^{20}\,{\rm m^{-3}}$ and we assume a constant plasma density
of $10^{20}\,{\rm m^{-3}}$ during these simulations. The $Z_{{\rm eff}}$
is taken to be $\sim1$.

In these simulations, the parameter regime is dictated by the value
of variable $a$ in Eq.(16), which is determined from the experimental
parameter to be $\simeq0.506$ and $0.285$ in SI units, for the shots
(\emph{a}) and (\emph{b}). The values of the index $\alpha$ are chosen
so as to get the desired relaxation oscillation with a given growth
rate $\rho$. In the cases shown in Fig.11, the values of $\alpha$
are taken to be 0.42 and 1.1, respectively with the dimensionless
growth rate $\rho=2000$, in both cases.

\begin{figure}
\begin{center}\includegraphics[%
  scale=1.2]{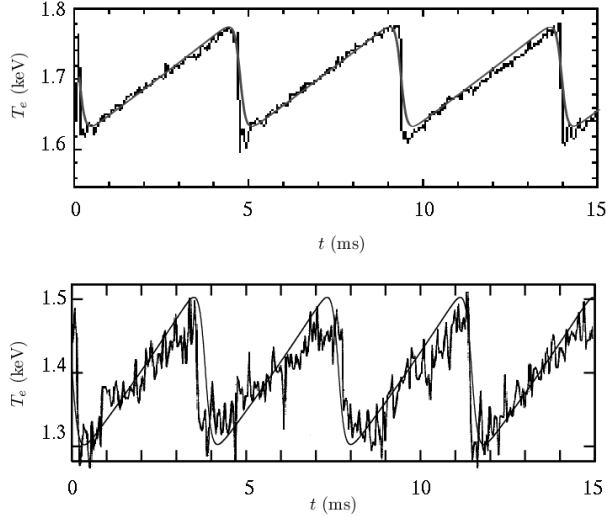}\end{center}

\caption{Approximate representation of two sets of experimental sawtooth oscillations.
The numerical solutions of Eqs.(16) and (17) (thick solid lines) are
superimposed on the actual experimental data (the zig-zag lines).}
\end{figure}

\section{The invariant manifold}

In this section, we construct the invariant manifold of Eqs.(16) and
(17), using the renormalization-group (RG) method \cite{kunihiro}
and show that it is indeed the slow manifold which governs the linear
rise of the electron temperature in the sawtooth on the limit cycle.
It has been shown elsewhere that the RG method can be applied to continue
the unperturbative solutions of evolution equations valid locally
around an arbitrary initial value $t=t_{0}$, smoothly to an arbitrary
time $t$ constructing the envelope of the perturbative solutions
\cite{shin}.

With the rescaling of the variables, we write Eqs.(16,17) as in Eqs.(45,46)\begin{eqnarray}
\dot{x} & = & x^{2}(1-ax^{2})-xy,\\
\dot{y} & = & \rho(x^{2}-1)y.\end{eqnarray}
 where we have rescaled the variables $t\rightarrow2t$ and $\rho\rightarrow\rho/2$.
Without loss of any generality we have assumed that $\alpha=1$. We
further make a scale transformation\begin{equation}
x=\epsilon\xi,\qquad y=\epsilon\eta,\end{equation}
 where $\epsilon$ is a small and arbitrary parameter. The evolution
equations now read\begin{eqnarray}
\dot{\xi} & = & \epsilon\xi^{2}(1-a\epsilon^{2}\xi^{2})-\epsilon\xi\eta,\\
\dot{\eta} & = & \rho(\epsilon^{2}\xi^{2}-1)\eta.\end{eqnarray}

We now try to construct a perturbative solution $\bu(t;t_{0})$ of
the above equations around some initial time $t=t_{0}$,\begin{eqnarray}
\left(\begin{array}{c}
\xi\\
\eta\end{array}\right)=\bu(t;t_{0}) & = & \bu_{0}(t;t_{0})+\epsilon\bu_{1}(t;t_{0})\nonumber \\
 &  & +\epsilon^{2}\bu_{2}(t;t_{0})+\mathcal{O}(\epsilon^{2}).\end{eqnarray}
 The differential equations at successive orders to be solved are
given as\begin{eqnarray}
(\partial_{t}-A)\bu_{0}(t;t_{0}) & = & 0,\\
(\partial_{t}-A)\bu_{1}(t;t_{0}) & = & (\xi_{0}^{2}-\xi_{0}\eta_{0})U_{2},\\
(\partial_{t}-A)\bu_{2}(t;t_{0}) & = & (2\xi_{0}\xi_{1}-\xi_{1}\eta_{0})U_{2}\nonumber \\
 &  & +\rho\xi_{0}^{2}\eta_{0}U_{1},\end{eqnarray}
 corresponding to the zeroth, first, and second orders. In the above
equations,\begin{equation}
A=\left(\begin{array}{cc}
0 & 0\\
0 & -\rho\end{array}\right)\end{equation}
 with eigenvalues $(-\rho,0)$ corresponding to the eigenvectors $U_{1}=(0,1)^{t}$
and $U_{2}=(1,0)^{t}$. The zeroth order solution is given by\begin{equation}
\bu_{0}(t;t_{0})=c_{1}(t_{0})U_{2}+c_{2}(t_{0})e^{-\rho t}U_{1}.\end{equation}
 The first and second order solutions are respectively\begin{eqnarray}
\bu_{1}(t;t_{0}) & = & \frac{c_{1}(t_{0})^{2}}{(\partial_{t}-A)}U_{2}-\frac{c_{1}(t_{0})c_{2}(t_{0})}{(\partial_{t}-A)}e^{-\rho t}U_{2},\\
 & = & c_{1}(t_{0})^{2}(t-t_{0})U_{2}\nonumber \\
 &  & +\frac{1}{\rho}c_{1}(t_{0})c_{2}(t_{0})(e^{-\rho t}-e^{-\rho t_{0}})U_{2},\\
\bu_{2}(t;t_{0}) & = & \int_{t_{0}}^{t}ds\,(2c_{1}(t_{0})-c_{2}(t_{0})e^{-\rho t})\nonumber \\
 &  & \times\left[\frac{}{}\! c_{1}(t_{0})(s-t_{0})\right.\nonumber \\
 &  & +\left.\frac{1}{\rho}c_{1}(t_{0})c_{2}(t_{0})(e^{-\rho t}-e^{-\rho t_{0}})\right]U_{2}.\end{eqnarray}
 Finally, the RG equation $\partial\bu(t;t_{0})/\partial t_{0}|_{t_{0}=t}=0$
yields\begin{eqnarray}
\dot{c}_{1}(t)-\epsilon c_{1}(t)\left(c_{1}(t)-c_{2}(t)e^{-\rho t}\right) & = & 0,\\
\dot{c}_{2}(t)-\epsilon^{2}\rho c_{1}(t)c_{2}(t) & = & 0.\end{eqnarray}
 Note that we have retained terms only up to the first order in the
first of the above equations. On the limit cycle, we solve the above
equations, Eqs.(79,80) for $c_{1,2}(t)$ and with $t\rightarrow\infty$.
Thus, from Eq.(70), the invariant manifold can be constructed as,\begin{equation}
\xi(t;t)=\frac{1}{C_{1}-\epsilon t},\qquad\eta(t;t)=0,\end{equation}
 with $C_{1,2}$ as constants of integration. So, this invariant manifold
in terms of variable $x$ and $y$ in Eqs.(16) and (17) is given by,\begin{equation}
x\sim\frac{\epsilon^{2}}{C_{1}^{2}}\left(1+\frac{2\epsilon t}{C_{1}}\right),\qquad y\sim0,\end{equation}
which essentially is the slow manifold given by Eqs.(54) and (55)
in Sec.IV.

\section{Conclusion}

To conclude, we have proposed a minimal dynamical model for the sawtooth
oscillations in current carrying plasmas e.g. tokamak plasmas. This
model, based on the assumption that the sawtooth is triggered due
to a thermal instability which ejects the plasma, shows relaxation
oscillations on an isolated limit cycle. We have further shown that
the invariant manifold of this model is indeed the slow manifold of
the relaxation oscillation. The persistent behaviour of the sawtooth
oscillation across different tokamaks indicate that a dynamical model
based on limit cycle oscillation is consistent in contrast to the
Hamiltonian models.

We note that, the Hopf bifurcation, which is the key to the limit
cycle behavior, around an unstable equilibrium point is based on a
linear theory, which does not always predict the behaviour of the
corresponding nonlinear model. However, in this case, numerical simulations
seem to be in agreement with the predicted bifurcation analysis.

\begin{ack}
The authors would like to thank the anonymous referees for suggesting
several improvements including addition of subsection covering the
stability analysis of the limit cycle. 
\end{ack}

\section*{Appendix}

\subsection*{Hamiltonian formalism}

In this Appendix, we review the equivalent Hamiltonian approximation
of Eqs.(16) and (17) \cite{coppi2,coppi3}. As can be shown that,
a Hamiltonian for Eqs.(16) and (17) exist only for $a=0$ and $\alpha=3/2$,
and the equations become \begin{eqnarray}
\dot{x} & = & x^{(3/2)}(1-y),\\
\dot{y} & = & \rho(x-1)y.\end{eqnarray}
In this case, the Hamiltonian of the system can be expressed as\begin{equation}
\mathcal{H}=\int^{p}\rho\left(\frac{4}{p^{2}}-1\right)\, dp+\int^{q}(e^{q}-1)\, dq,\end{equation}
with the equivalent Hamilton's equations as\begin{equation}
\frac{dq}{dt}=\frac{\partial\mathcal{H}}{\partial p},\qquad\frac{dp}{dt}=-\frac{\partial\mathcal{H}}{\partial q}.\end{equation}
The equivalent conjugate variables i.e `position' and `momentum' are,
respectively, given by,\begin{equation}
q=\ln y,\qquad p=-2/\sqrt{{x}}.\end{equation}
The Hamiltonian formalism ensures periodic orbits described by the
family of curves described by the differential equation,\begin{equation}
\frac{dy}{dx}=\frac{\rho(x-1)y}{x^{3/2}(1-y)},\end{equation}
which are closed. However, depending on the initial conditions of
evolution, there can be several possible orbits rather than an unique
one. Note that the Eqs.(83,84) are similar in nature to the famous
Volterra's predator-prey system \cite{strogatz}.

\end{document}